\documentclass[aps,showpacs,preprintnumbers,amsmath,amssymb,prl,twocolumn
]{revtex4-1}
\usepackage{txfonts}
\usepackage{graphicx}
\usepackage{dcolumn}
\usepackage{bm}
\usepackage{color}
\def\comment#1{}

\def\slashchar#1{\setbox0=\hbox{$#1$}           
   \dimen0=\wd0                                 
   \setbox1=\hbox{/} \dimen1=\wd1               
   \ifdim\dimen0>\dimen1                        
      \rlap{\hbox to \dimen0{\hfil/\hfil}}      
      #1                                        
   \else                                        
      \rlap{\hbox to \dimen1{\hfil$#1$\hfil}}   
      /                                         
   \fi}                                         %

\def\nablab{{\mbox{\boldmath $\nabla$}}}

\begin{document}

\title{Fluctuation-induced magnetization dynamics and criticality at the interface of a
topological insulator with a magnetically ordered layer}

\author{Flavio S. Nogueira}
\author{Ilya Eremin}
\affiliation{Institut f{\"u}r Theoretische Physik III, Ruhr-Universit\"at Bochum,
Universit\"atsstra\ss e 150, 44801 Bochum, Germany}

\date{Received \today}

\begin{abstract}
We consider  a theory for a two-dimensional interacting conduction electron system with strong spin-orbit coupling on the interface between a topological insulator and the magnetic (ferromagnetic or antiferromagnetic) layer.
For the ferromagnetic case we derive the Landau-Lifshitz equation, which features a contribution proportional to
a fluctuation-induced electric field obtained by computing the topological (Chern-Simons)
contribution from the vacuum polarization. We also show that fermionic quantum fluctuations reduce the critical
temperature $\tilde T_c$ at the interface relative to the critical temperature $T_c$ of the bulk, so that
in the interval $\tilde T_c\leq T<T_c$ is possible to have coexistence of gapless Dirac fermions at the interface
with a ferromagnetically ordered layer.
For the case of an antiferromagnetic layer on a topological insulator substrate,
we show that a second-order quantum phase transition occurs at the interface, and compute
the corresponding critical exponents. In particular, we show that the electrons at the interface acquire an
anomalous dimension at criticality. The critical behavior of the N\'eel order parameter is anisotropic and
features large anomalous dimensions for both the longitudinal and transversal fluctuations.
\end{abstract}

\pacs{75.70.-i,73.43.Nq,64.70.Tg,75.30.Gw}
\maketitle
A spin current may exhibit interesting topological properties in systems where a Berry curvature in
Bloch momentum space is
induced by the underlying band structure \cite{Zhang-2003,Zhang-2004}, like for example in the case
of some hole-doped semi-conductors described by a Luttinger Hamiltonian \cite{Luttinger-1956} or systems featuring a
Rashba spin-orbit coupling \cite{Sinova-PRL-2004}. More recent prominent examples involve the quantum spin Hall
insulators or topological insulators (TI) \cite{Hasan-Kane-RMP,Zhang-RMP-2011},
where a Berry curvature in momentum space also arises. Depending on
the physical situation the Berry curvature may be Abelian or non-Abelian, and determines a magnetic monopole in
momentum space.

The surface of a TI, when in contact with a material exhibiting magnetic order, offers 
a framework for many topological effects. For instance, the two-dimensional system represented by the surface of
a topological insulator can be used as substrate for a magnetic layer, which can be either
ferromagnetic (FM) or antiferromagnetic (AF).
For a FM layer having a TI as substrate, a theoretical study of the magnetization dynamics was
carried out recently \cite{Nagaosa-2010}. In a similar context, the electric charging of magnetic textures has also
been discussed \cite{Nagaosa-2010-1}. Other interesting electromagnetic topological effects with a similar setup were
studied \cite{Franz-2010,Qi-2008,Rosenberg-2010,Essin-2010} and have shown to exhibit properties
similar to those of axion electrodynamics \cite{Axions}. In the axion electrodynamics a
topological term of the form
$(8\pi^2)^{-1}\alpha\theta{\bf E}\cdot{\bf B}$ is present \cite{Qi-2008,Axions} in the action, where $\alpha$ is the fine structure constant
and $\theta$ is the so called axion field. For the case where $\theta$
is uniform, time-reversal invariant TI's require $\theta=\pi$ \cite{Qi-2008}. Such a term should play a very important role at interfaces of TIs with other insulators. For a magnetic insulating layer on the surface of a TI,
a modification of the magnetization dynamics occurs, due to a direct coupling of the magnetization to the electric field.
Indeed, we have ${\bf E}\cdot{\bf B}={\bf E}\cdot({\bf H}+4\pi{\bf M})$, giving rise to a magnetoelectric effect,
which influences the precession of the magnetization \cite{Franz-2010}.

At the same time, the experimental situation is far from being clear. 
For instance, from a theory perspective one would expect that the coupling of a TI to a FM layer makes the surface states gapped.
However, in a very recent experiment
\cite{Rader-2012} where Fe impurities were deposited on Bi$_2$Se$_3$, no sign for a gap has been found,
in apparent conflict with theoretical expectations. Therefore,
further theoretical studies on the coupling of a TI substrate to magnetic system are necessary.

In this paper we consider quantum fluctuations effects stemming from the proximity-induced magnetism on the
surface of a TI. We assume that the electrons on the surface of the TI interact via
a long-range Coulomb interaction. For the case of a  FM layer in contact with the TI,
we will derive a Landau-Lifshitz (LL) equation which accounts for these interaction effects.
In our calculation an axion-like term emerges due to quantum fluctuations.
At the interface, it manifests itself as
a Chern-Simons (CS) term \cite{CS}, which breaks time-reversal symmetry, as a consequence of the coupling
of the surface of the TI to the magnetic layer. Furthermore, the electronic quantum fluctuations make the
stiffness anisotropic, even if the bulk of the FM layer features an isotropic stiffness. We also show that
due to the quantum fluctuations of the electrons, the critical temperature $\tilde T_c$
at the interface is reduced relative to the critical temperature $T_c$ of the FM layer. This
allows the existence of gapless fermions at the interface at the temperature range $\tilde T_c\leq T<T_c$ where
the bulk magnetic layer is still magnetically ordered.

It has been shown in a recent study \cite{Qi-2012} that the best candidate material to gap the topological surface states of a TI is 
MnSe, which is an AF insulator. 
For this case we will show that a second-order quantum phase transition occurs at the interface, and that it defines
a new universality class. One consequence of this interface quantum criticality is that the surface electrons become
gapless at the quantum critical point (QCP). This does not happen 
in the FM case we study. 
Hence, the topologically protected gapless modes can be restored at zero temperature by disordering the
AF long-range order at the interface.
A further important feature of this interface quantum criticality is the emergence of a large anomalous dimension
for the N\'eel order parameter. Interestingly, at the QCP the fermions will also acquire an anomalous dimension.

Our starting point is the Lagrangian for conduction electrons interacting via a Coulomb interaction
on the surface of an insulator either in contact with a bulk FM composed of several layers,
similarly to Ref. \cite{Nagaosa-2010}, or with an AF bulk system. Thus, if ${\bf n}$ is the induced magnetization 
at the interface and ${\bf L}$ 
the angular momentum, the spin of the conduction electrons, ${\bf S}=(1/2)c^\dagger\vec \sigma c$, is coupled to the 
total magnetization $(\mu_B/2){\bf L}+{\bf n}$ via an exchange term $-2J{\bf S}\cdot[(\mu_B/2){\bf L}+{\bf n}]$, 
where 
$c^\dagger=[c_\uparrow^\dagger~~c_\downarrow^\dagger]$, with $\vec \sigma=(\sigma_x,\sigma_y,\sigma_z)$
being the Pauli matrices. 
The lack of inversion symmetry in the direction perpendicular to the interface leads to 
a spin-orbit coupling of the Rashba type. Thus, the Lagrangian for the conduction electrons at the interface reads 
(we are assuming units where $\hbar=c=1$), 
\begin{equation}
\label{Eq:Lc}
 {\cal L}_c=c^\dagger[i\partial_t+e\varphi -iv(\sigma_y\partial_x-\sigma_x\partial_y)+J\vec \sigma\cdot{\bf n}]c
 -\frac{\epsilon}{4\pi}\varphi|\nablab|\varphi,
\end{equation}
where $v\propto J$. We will give further details on the exchange part of the Lagrangian shortly.       
In writing the above
Lagrangian we have assumed that the spin-orbit coupling is much stronger than the usual kinetic term of
the conduction electrons, which has been neglected. The auxiliary (Hubbard-Stratonovich)
field $\varphi$ accounts for the Coulomb interaction. Upon
integrating out $\varphi$ the usual Coulomb interaction between the electrons is obtained. The non-local Gaussian term
for $\varphi$ reflects the three-dimensional character of the Coulomb interaction in a two-dimensional problem,
similarly to graphene \cite{interac-graphene}. In this term $\nablab$ is the two-dimensional gradient and 
$\epsilon$ represents the dielectric constant.

The full Lagrangian of the systems includes the Lagrangian
describing the magnetization dynamics of the bulk FM, 
which includes a Landau-Ginzburg (LG) type functional and is given by
\begin{equation}
 {\cal L}_{\rm FM}={\bf b}\cdot\partial_t{\bf n}-\frac{\kappa}{2}[(\nablab{\bf n})^2+(\partial_z{\bf n})^2]-\frac{m^2}{2}{\bf n}^2-\frac{u}{4!}({\bf n}^2)^2,
\end{equation}
where $\kappa,u>0$ and $m^2=a_0(T-T_0)$,  
with $T_0$ being the (mean-field) critical temperature to disorder the FM. 
${\bf b}$ is the Berry connection, which fulfills the
usual monopole condition,
$\partial b_i/\partial n_j-\partial b_j/\partial n_i=\epsilon_{ijk}n_k/{\bf n}^2$.
For $m^2<0$ (or $T<T_0$) the bulk FM  is in a ferromagnetically ordered state.

Before considering the magnetization dynamics, 
let us first consider a fluctuation-corrected mean-field theory where the only 
fluctuation effects that are taken into account are the fermionic ones, i.e., 
$\sigma$ is assumed to be uniform, and the transverse
fluctuations of the magnetization vanish. We also neglect the fluctuation effects of the Coulomb interaction. 
The calculations are  done in imaginary time and finite temperature. 

In this case, after integrating out the 
fermions, we obtain the free energy density,
\begin{equation}
 {\cal F}=-T\sum_{n=-\infty}^\infty\int\frac{d^2p}{(2\pi)^2}\ln(\omega_n^2+v^2{\bf p}^2+J^2\sigma^2)+\frac{m^2}{2}\sigma^2+\frac{u}{4!}\sigma^4,
\end{equation}
where $\omega_n=(2n+1)\pi T$ is the fermionic Matsubara frequency. After performing the Matsubara sum, the 
remaining integral over momenta contains a zero temperature contribution which is divergent, requiring 
regularization and renormalization. Using an ultraviolet cutoff $\Lambda\sim a^{-1}$, where $a$ is the lattice constant, we can cancel 
the dependence on the cutoff by minimally absorbing it in a redefinition of the Curie temperature of the bulk FM 
precisely at the interface. The physical 
requirement (or renormalization condition) is that the zero temperature fermionic gap, $m_\psi\equiv J\sigma_0$, 
is finite in the long-wavelength limit. 

The saddle-point approximation yields,
\begin{equation}
\label{Eq:gap-eq}
 a_0(T_c-T)=\frac{u}{6}\sigma^2+\frac{J^2T}{\pi v^2}\ln\left[2\cosh\left(\frac{J\sigma}{2T}\right)\right],
\end{equation}
where $T_c$ is the renormalized Curie temperature of the bulk FM at the interface. 
The critical temperature, $\tilde T_c$, at the interface is obtained by 
demanding that $\sigma$ vanishes at $T=\tilde T_c$. This yields $\tilde T_c=T_c[1+J^2\ln 2/(\pi a_0v^2)]^{-1}$. On the other hand, 
by setting $T=0$ in Eq. (\ref{Eq:gap-eq}), we obtain 
that {\it at the interface} $T_c=um_\psi^2/(6J^2)+J^2m_\psi/(2\pi a_0v^2)$. Note that this expression makes only sense at the interface and does 
not correspond to the physical critical temperature there, which is actually given by $\tilde T_c$. 
Furthermore, since $v\propto J$ and at leading order $\sigma_0^2\approx 6a_0T_0/u$, we obtain that $T_c\to T_0$ as $v\to 0$. 
In order to estimate $\tilde T_c$, we assume that $T_c\gg um_\psi^2/(6J^2a_0)$, such that we have approximately, 
$\tilde T_c\approx m_\psi T_c[m_\psi+(2\ln 2)T_c]^{-1}$. If we use the estimates $m_\psi\approx 28.2$ meV and 
$T_c\approx 70$ K \cite{Qi-2012,Wan}, we obtain $\tilde T_c\approx 54$ K.   
Thus, our fluctuation-corrected mean-field theory implies that the critical temperature at the interface is smaller than the Curie temperature 
of the bulk FM. Therefore, it is possible to destroy the proximity-induced magnetization at the interface while the bulk FM is 
still ordered. This occurs typically in a temperature window $\tilde T_c\leq T<T_c$, where we are assuming that $T_c$ does not differ 
appreciably from $T_0$. 
The reduction of the critical temperature at the interface with respect to the bulk one is 
an important consequence of the interplay between ferromagnetic proximity effect and the spin-orbit 
coupling.

\begin{figure}
 \includegraphics[width=7cm]{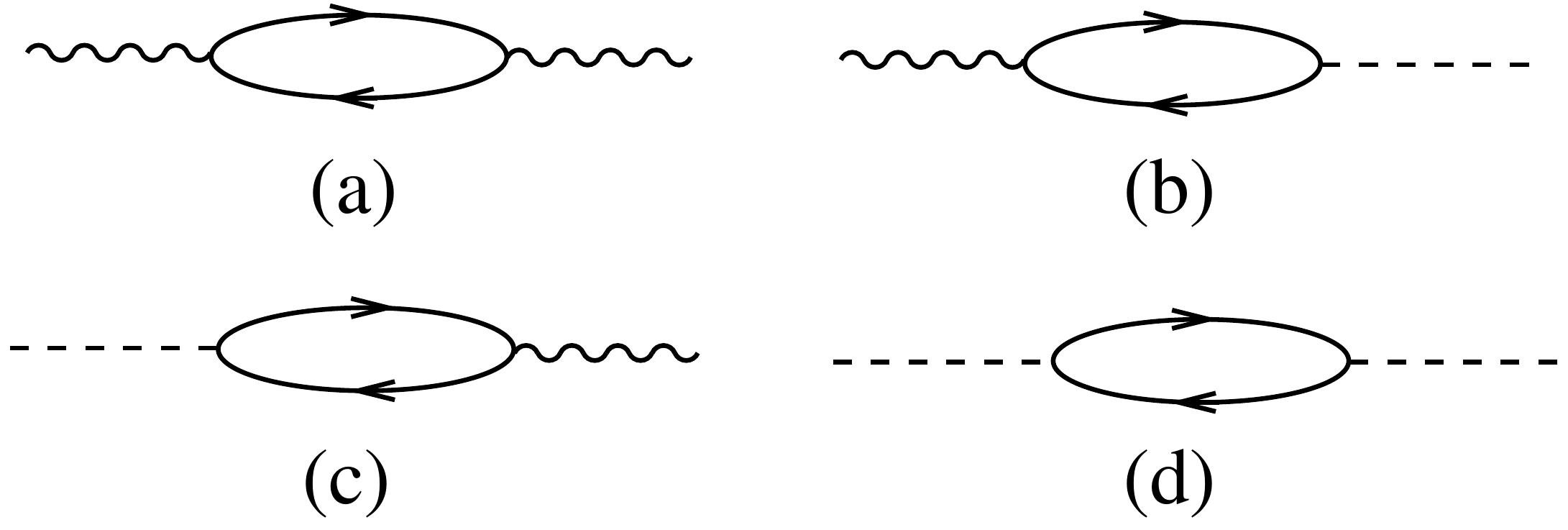}
 \caption{Diagrams contributing to $\delta S_{\rm gauge}$, Eq. (\ref{Eq:S-gauge}). The wiggled line represents
 the vector field $a_\mu$, the solid line is a Dirac fermion, and the dashed line represents the
 fluctuating part of the $\sigma$ field. The diagrams (b) and (c) cancel out.}
 \label{Fig:diagrams}
\end{figure}

The next step is to compute the fluctuations of the order parameter around the mean-field theory.
Since we are interested in deriving a differential equation for the magnetization dynamics, we will
return to real time in the following and consider a zero temperature calculation. 
In order to facilitate our analysis of the problem, it is convenient to rewrite the Lagrangian for the
conduction electrons in a QED-like form, which is achieved by the rescalings $\varphi\to (J/e)\varphi$,
$x_i\to vx_i$ ($i=1,2$), in the action, to obtain
\begin{equation}
{\cal L}_c=\bar \psi(i\slashchar{\partial}-J\slashchar{a})\psi+J\sigma\bar \psi\psi
 -(\zeta/2)a_0|\nablab|a_0,
 \end{equation}
where the Dirac matrices are defined by $\gamma^0=\sigma_z$, $\gamma^1=-i\sigma_x$, and $\gamma^2=i\sigma_y$,
$\psi=vc$, such that the usual relativistic notations for spinors hold with
$\bar \psi=\psi^\dagger\gamma^0$, and also the usual Dirac slash notation, $\slashchar{Q}=\gamma^\mu Q_\mu$, is being used.
The gauge field is given by $a^\mu=(\varphi,n_y,-n_x)$ and
$\sigma=n_z$, and the dielectric constant, $\zeta\equiv \epsilon vJ^2/(2\pi e^2)$. 
We will  assume that there are $N$ fermionic orbital degrees of freedom.
Thus, integrating out the fermions yields the gauge-invariant contribution to the effective action,
$S_{\rm gauge}=iN{\rm Tr}\ln(i\slashchar{\partial}-J\slashchar{a}+J\sigma)$.

The lowest order diagrams associated to the fluctuating fields are shown in Fig. \ref{Fig:diagrams}.
Approximate evaluation of $S_{\rm gauge}$ at long wavelengths yields the leading fluctuation contribution,
$S_{\rm gauge}\approx S_{\rm eff}^{\rm MF}+\delta S_{\rm gauge}$, where
\begin{eqnarray}
\label{Eq:S-gauge}
 \delta S_{\rm gauge}&=&\frac{NJ^2}{8\pi}\int dt\int d^2r\left\{-\frac{1}{3m_\psi}
 (\epsilon_{\mu\nu\lambda}\partial^\nu a^\lambda)^2
 \right.\nonumber\\
 &+&\left.\epsilon_{\mu\nu\lambda}a^\mu\partial^\nu a^\lambda
 +\frac{1}{m_\psi}[(\partial_t\tilde \sigma)^2-(\nablab\tilde \sigma)^2]\right\},
\end{eqnarray}
with the fluctuation $\tilde \sigma$ arising from the decomposition $\sigma=\sigma_0+\tilde \sigma$.
The quadratic fluctuation term in $\tilde \sigma$ will generate an anisotropy in the magnetic system, which
is isotropic in the bulk.
The first term in Eq. (\ref{Eq:S-gauge}) corresponds to a Maxwell term in $(2+1)$-dimensional electrodynamics.
The second term is a CS term \cite{CS} generated
by the quantum fluctuations. This CS term reflects the breaking of time-reversal symmetry due to the coupling to a magnetic
layer. In order to better appreciate the effect of the CS term, it is useful to rewrite the CS contribution to $S_{\rm gauge}$
in the form,
\begin{equation}
\label{Eq:S-CS}
 S_{\rm CS}=\frac{\sigma_{xy}}{4\pi}\int dt\int d^2r(n_y\partial_tn_x-n_x\partial_tn_y+2{\bf n}\cdot\nablab\varphi),
\end{equation}
where $\sigma_{xy}=\sigma^0_{xy}NJ^2/e^2$, with $\sigma^0_{xy}=e^2/2$ (in units where $\hbar=1$), is the induced Hall conductivity.
It is readily seen that the contribution proportional
to $n_x\partial_tn_y-n_y\partial_tn_x$ yields an additional Berry phase, as discussed previously in Ref. \cite{Nagaosa-2010}.
The term proportional to ${\bf n}\cdot\nablab\varphi$ is a crucial contribution stemming from the Coulomb interaction between
the fermions at the interface. Indeed,
since $\varphi$ is a fluctuating scalar potential associated to the Coulomb interaction, this term yields a contribution
proportional to ${\bf M}\cdot{\bf E}$, where ${\bf E}=-\nablab\varphi$ is a fluctuation-induced electric field,
and ${\bf M}\sim{\bf n}$.
Thus, this term corresponds to an emergent axion-like term.

The Landau-Lifshitz (LL) equation for the magnetization dynamics at the interface (i.e., at $z=0$)
can now be obtained from the Euler-Lagrange equation for the effective action. 
We have, 
\begin{eqnarray}
 \label{Eq:LL}
\partial_t{\bf n}&=&{\bf n}\times\left\{\rho_s^{ij}(\nabla^2{\bf n})_j{\bf e}_i
+\frac{Z\sigma_{xy}}{2\pi v^2m_\psi}[(\partial_t^2{\bf n})_z{\bf e}_z-\nablab(\nablab\cdot{\bf n})]\right\}
\nonumber\\
&+&\frac{Z\sigma_{xy}}{2\pi v^2}\left[{\bf n}\times{\bf E}+\frac{1}{3m_\psi}({\bf n}\cdot{\bf e}_z)\partial_t{\bf E}\right],
\end{eqnarray}
where the stiffness matrix elements are given by  
$\rho_s^{ij}=(Z/v^2)[\kappa(\delta_{ix}\delta_{jx}+\delta_{iy}\delta_{jy})+(\kappa+\sigma_{xy}/(2\pi m_\psi))\delta_{iz}\delta_{jz}]$,
with $Z=[1-m_\psi\sigma_{xy}/(2\pi v^2 J)]^{-1}$. For the LL equation in the bulk one has to supply the boundary conditions 
for the bulk magnetization, which must reflect the influence of the surface states of the TI over some penetration depth into the 
bulk FM, assumed to be semi-infinite, having a surface at $z=0$ coinciding with the interface with the TI. The relevant boundary 
conditions at $t=0$ are,
 $\partial_z{\bf n}|_{z=0}=-J\langle c^\dagger\vec \sigma c\rangle$,
$\partial_z{\bf n}|_{z=\infty}=0$, and $\lim_{z\to\infty}{\bf n}({\bf r},z)={\bf n}_b({\bf r})$, where ${\bf n}_b({\bf r})$ is the 
bulk magnetization far away from the interface. Let us see how these boundary conditions work within a simple 
mean-field approximation at $T=0$ and consider the following Ansatz for the magnetization precession in the bulk FM,   
${\bf n}({\bf r},z,t)=2^{-1/2}\sigma(z)[\cos({\bf k}\cdot{\bf r}-\omega t){\bf e}_x
+\sin({\bf k}\cdot{\bf r}-\omega t){\bf e}_y+{\bf e}_z]$, where $\omega\propto k^2$. Thus, we have ${\bf n}^2=\sigma^2(z)$. 
The boundary conditions at the interface are $\sigma(\infty)=\sigma_b$,     
$\partial_z\sigma|_{z=0}=-J^3\sigma_0^2/(2\pi v^2)$, where $\sigma_0=\sigma(0)=m_\psi/J$. We will define 
the $k$-dependent length characterizing the longitudinal magnetization in the bulk, 
$\xi_b(k)=(a_0T_0/\kappa-k^2)^{-1/2}$, where $k^2<a_0T_0/\kappa$.   
The magnetization $\sigma$ can now be determined \cite{Mills} 
by solving {\it exactly} the equation $\partial^2\sigma/\partial z^2+\xi_b^{-2}\sigma-(u/6\kappa)\sigma^3=0$.  
We obtain $\sigma(z)=\sigma_b(1-\Delta\sigma e^{-\sqrt{2}z/\xi_b})^{-1}(1+\Delta\sigma e^{-\sqrt{2}z/\xi_b})$, 
where $\Delta\sigma=(\sigma_0-\sigma_b)/(\sigma_0+\sigma_b)$. The boundary condition at $z=0$ yields  
$8\pi v^2\sigma_b\Delta\sigma=2^{-1/2}J^3\xi_b(1-\Delta\sigma)^2\sigma_0^2$, which determines $\xi_b$ 
in terms of $\sigma_0$ and $\sigma_b$. Note that this condition yields $\sigma_0=\sigma_b$ for $J=0$, as it should. 
This calculation shows that electrons on the surface of the TI influence the magnetization dynamics of the bulk over 
a characteristic length $\sim\xi_b$ which is uniquely determined by the boundary conditions.

Eq. (\ref{Eq:LL}) is one of the main results of this paper. It
leads to a fluctuation-induced magnetoelectric effect. 
One important consequence of Eq. (\ref{Eq:LL}) 
is that due to the fluctuation-induced electric field, the magnitude of 
the magnetization is not constant, as it would be in the case of absence of Coulomb interaction or for a constant electric field. 
In particular, if the electric field is only due to external effects, this result implies that  
we can use a time-dependent electric field to control the magnitude of the magnetization.    

Part of the coupling to the electric field, discussed
previously by Garate and Franz \cite{Franz-2010}, is reproduced here as a fluctuation effect due to the Coulomb interaction
between the spin-orbit coupled electrons lying on the surface of a TI. We have obtained in addition a contribution involving  
$\partial_t{\bf E}$ that accounts for the time-dependence of the electric field. Note that the term involving 
$(\partial_t^2{\bf n})_z$ is typically small at low energy and can be safely neglected in most calculations. 

Next we discuss the case of an AF layer on a TI substrate at zero temperature, which, as we will see, differs fundamentally from the case of
a FM layer.
In the AF case  a quantum phase transition occurs at the interface. In order to study the
phase structure of the theory in this case, we will work only in imaginary time from now
on. Specifically, we consider an Euclidean effective field theory whose Lagrangian has the form,
\begin{eqnarray}
\label{Eq:L-eff-AF}
{\cal L}&=& \bar \psi(\slashchar{\partial}-ig_1\slashchar{a}+g_2\sigma)\psi+\frac{\zeta}{2}a_0|\nablab|a_0
+\frac{1}{2}[(\partial_\mu\sigma)^2+(\partial_\mu{\bf a})^2]
\nonumber\\
&+&\frac{M^2}{2}(\sigma^2+{\bf a}^2)
+\frac{\lambda}{4!}(\sigma^2+{\bf a}^2)^2,
\end{eqnarray}
where we are not making any longer distinction between upper and lower covariant indices, since the metric of the
theory has now Euclidean signature. Note that we are also assuming that $g_1\neq g_2$, as quantum fluctuations 
induces an anisotropy.  
In the spirit of effective field theories,
the coupling constants are understood as effective parameters to be determined by
a renormalization group (RG) flow. Thus, the phase structure of
the theory is completely determined by the RG equations for the coupling constants.

At low energies and one-loop order (see below)
the fixed point structure will be governed by the dimensionless couplings  
$\hat g_i^2=g_{i,r}^2/M_r^\epsilon$ ($i=1,2$) and $\hat \lambda=\lambda_r/M_r^\epsilon$, where $\lambda_r$ and $g_{i,r}$ are corresponding
renormalized couplings and we are using the renormalized mass $M_r$ as renormalization scale \cite{ZJ}. Here
$\epsilon=4-d$, where $d=D+1$, with $D$ being the spatial dimension. Our analysis is done in the framework of
the $\epsilon$-expansion, which is carried out up to one-loop order.
As usual, in such a renormalization scheme
the renormalized mass gives the inverse of the correlation length, i.e., $M_r=\xi^{-1}$.
Due to the coupling between $\sigma$ and the fermions, a mass anisotropy will be generated, defining in this way two correlation lengths,
related to longitudinal and transversal fluctuations. We will assume that $\xi$ refers to the longitudinal correlation length, giving
the fluctuations of the $\sigma$ field. The correlation length due to transversal fluctuations will be denoted by $\xi_\perp$. If
$\nu$ and $\nu_\perp$ are respectively the critical exponents of the longitudinal and transversal correlation lengths, we
easily obtain that $\xi\sim\xi^{\nu_\perp/\nu}$, which determines the crossover exponent $\phi=\nu/\nu_\perp$.
The quantum critical behavior
can be derived from a generalization of the extended Gross-Neveu model \cite{GN} discussed in Ref. \cite{ZJ-NPB-1991}.
We obtain in this way
the one-loop $\beta$ functions $ \beta_{\hat g_1^2}\equiv M_r\partial \hat g_1^2/\partial M_r$,
$\beta_{\hat g_2^2}\equiv M_r\partial \hat g_2^2/\partial M_r$ and
$\beta_{\hat \lambda}\equiv M_r\partial\hat \lambda/\partial M_r$ in the form,
%
 $\beta_{\hat g_1^2}=-\epsilon\hat g_1^2+N\hat g_1^4/(12\pi^2)$, 
%
 $\beta_{\hat g_2^2}=-\epsilon\hat g_2^2+(N+3)\hat g_2^4/(8\pi^2)$, and 
%
 $\beta_{\hat \lambda}=-\epsilon\hat u+(8\pi^2)^{-1}[(11/2)\hat \lambda^2+2N\hat \lambda\hat g_2^2-12N\hat g_2^4]$.
The $\beta$ function for $\hat \zeta=\zeta_r/M_r$ follows from the non-locality of the quadratic term in $a_0$. Since
counterterms are local, this term does not renormalize, which implies simply $\beta_{\hat \zeta}=(N\hat g_1^2/12\pi^2-\epsilon)\hat \zeta$.

The quantum critical point is determined by demanding that the $\beta$ functions vanish, which yields the
infrared stable fixed points, $\hat g_{1*}^2=12\pi^2\epsilon/N$, $\hat g_{2*}^2=8\pi^2\epsilon/(N+3)$, and 
$\hat \lambda_*=8\pi^2\epsilon\left(3-N+\sqrt{N^2+258N+9}\right)/[11(N+3)]$.
The anomalous dimension $\eta_N$ of the N\'eel order parameter at the interface can be defined via
the scaling behavior  $\langle \sigma\rangle\sim M_r^{(2-\epsilon+\eta_N)/2}$, and is given at one-loop by
 $\eta_N=N\hat g_{2*}^2/(8\pi^2)=N\epsilon/(N+3)$.
For $N=1$ and two spatial dimensions (corresponding to $\epsilon=1$), we obtain $\eta_N=1/4$. This large
value of the anomalous dimension, as compared to the value obtained from the $O(3)$ universality class,
reflects the fact that $\langle\sigma\rangle$ receives contributions from
the composite operator $\bar \psi\psi$. The scaling behavior of ${\bf n}=(n_x,n_y,\sigma)$ at the interface is anisotropic,
and the transversal fluctuations have a different anomalous dimension, which is dominantly determined by the
vacuum polarization diagrams, $\eta_N^\perp=\epsilon$, yielding $\eta_N^\perp=1$ for $D=2$.
It is worth to mention that two-loop corrections will be small, but positive (typically $\sim 0.03$).
Therefore, we expect that a more accurate value for the anomalous dimension $\eta_N^\perp$ is slightly above the unity.

The electrons at the interface also have an anomalous scaling at the quantum critical point. 
This is in contrast with the FM case, where the fermionic spectrum
is always gapped at zero temperature.
Thus, we obtain the low-energy behavior, $\langle\bar \psi(p)\psi(p)\rangle\sim-i\slashchar{p}/p^{2-\eta_\psi}$,
%
where
%
 $\eta_\psi=\hat g_{2*}^2/(16\pi^2)=\epsilon/[2(N+3)]$.
For $D=2$ and $N=1$, we obtain $\eta_\psi=1/8$. Note that $\eta_\psi$ does not receive any contribution from
the fixed point $\hat g_{1,*}^2$ at one loop order. This is due to the fact that the vector field propagator takes
here the same form as one in QED where the Feynman gauge has been fixed.

It remains to compute the critical exponents of the correlation lengths.
The longitudinal correlation length exponent is given by $\nu=(2+\eta_M)^{-1}$, where at one-loop,
$ \eta_M=-5\hat \lambda_*/(48\pi^2)-\eta_N$.
%
Thus, by expanding up to first order in $\epsilon$, we obtain,
%
 $\nu\approx 1/2+\epsilon[4(N+3)]^{-1}[(5/66)(3-N+\sqrt{N^2+258N+9})+N]$.
Setting once more $D=2$ and $N=1$, we obtain $\nu\approx 0.649$. The transversal correlation length exponent, on the other hand,
is given by $\nu_\perp=(2+\eta_M^\perp)$, where $\eta_M^\perp=\eta_M-\eta_N^\perp+\eta_N$.
Thus, we obtain $\nu_\perp\approx\nu+3\epsilon/[4(N+3)]$,
which for $N=1$ and $D=2$ yields
$\nu_\perp\approx 0.83$. Note that the values of the correlation length exponents differ appreciably from the
one-loop value of the $O(3)$ universality class, $\nu_{O(3)}^{\rm one-loop}\approx 0.614$. 

In conclusion,  we have shown in the FM case that an axion-like term is generated
in the form of a CS term, which in turn modifies the magnetization dynamics of the LL equation.
Furthermore, we have shown that for a specific temperature window is possible to have gapless fermions at the interface and,
at the same time, a ferromagnetically ordered layer. 

For the case of an AF layer, we
have shown that a quantum phase transition occurs at the interface, and that the fermion spectrum becomes gapless at the QCP. Moreover,
large values of the anomalous dimensions for the N\'eel order parameter were obtained.

\acknowledgments

We would like to thank Karl Bennemann, Lars Fritz, and Dirk Morr for many interesting discussions. 
We would also like to thank the Deutsche Forschungsgemeinschaft (DFG) for the financial support via
the SFB TR 12.


\begin{thebibliography}{100}

\bibitem{Zhang-2003} S. Murakami, N. Nagaosa, and S.-C. Zhang, Science {\bf 301}, 1348 (2003).

\bibitem{Zhang-2004} S. Murakami, N. Nagaosa, and S.-C. Zhang, Phys. Rev. B {\bf 69}, 235206 (2004).

\bibitem{Luttinger-1956} J. M. Luttinger, Phys. Rev. {\bf 102}, 1030 (1956).

\bibitem{Sinova-PRL-2004} J. Sinova, D. Culcer, Q. Niu, N. A. Sinitsyn, T. Jungwirth, and A. H. MacDonald,
Phys. Rev. Lett. {\bf 92}, 126603 (2004).

\bibitem{Hasan-Kane-RMP} M. Z. Hasan and C. L. Kane, Rev. Mod. Phys. {\bf 82}, 3045 (2010).

\bibitem{Zhang-RMP-2011} X.-L. Qi and S.-C. Zhang, Rev. Mod. Phys. {\bf 83}, 1057 (2011).



\bibitem{Nagaosa-2010} T. Yokoyama, J. Zang, and N. Nagaosa, Phys. Rev. B {\bf 81}, 241410(R) (2010).

\bibitem{Nagaosa-2010-1} K. Nomura and N. Nagaosa, Phys. Rev. B {\bf 82}, 161401(R) (2010).

\bibitem{Franz-2010} I. Garate and M. Franz, Phys. Rev. Lett. {\bf 104}, 146802 (2010).

\bibitem{Qi-2008} X.-L. Qi, T. L. Hughes, and S.-C. Zhang, Phys. Rev. B {\bf 78}, 195424 (2008).

\bibitem{Rosenberg-2010} G. Rosenberg and M. Franz, Phys. Rev. B {\bf 82}, 035105 (2010).

\bibitem{Essin-2010} A. M. Essin, A. M. Turner, J. E. Moore, and D. Vanderbilt, Phys. Rev. B {\bf 81}, 205104 (2010).

\bibitem{Axions} E. Witten, Phys. Lett. B {\bf 86}, 283 (1979); F. Wilczek, Phys. Rev. Lett. {\bf 58}, 1799 (1987).

\bibitem{Rader-2012} M. R. Scholz, J. S\'anchez-Barriga, D. Marchenko, A. Varykhalov, A. Volykhov, L. V. Yashina, and
O. Rader, Phys. Rev. Lett. {\bf 108}, 256810 (2012).

\bibitem{CS} S. Deser, R. Jackiw, and S. Templeton, Phys. Rev. Lett. {\bf 48}, 975 (1982).

\bibitem{Qi-2012} W. Lao and X.-L. Qi, arXiv:1208.4638.




\bibitem{interac-graphene} I. F. Herbut, Phys. Rev. Lett. {\bf 97}, 146401 (2006); D. E. Sheehy and J. Schmalian,
Phys. Rev. Lett. {\bf 99}, 226803 (2007).

\bibitem{Wan} X. Wan, J. Dong, and S. Y. Savrasov, Phys. Rev. B {\bf 83}, 205201 (2011). 

\bibitem{Mills} D. J. Mills, Phys. Rev. B {\bf 3}, 3887 (1971); P. Kumar, Phys. Rev. B {\bf 10}, 2928 (1974).

\bibitem{ZJ} See Chapter 25 in: J. Zinn-Justin, {\it Quantum Field Theory
and Critical Phenomena}, 2nd ed. (Oxford University Press, Oxford, 1993).
\bibitem{GN} D. J. Gross and A. Neveu, Phys. Rev. D {\bf 10}, 3235 (1974).

\bibitem{ZJ-NPB-1991} J. Zinn-Justin, Nucl. Phys. B {\bf 367}, 105 (1991); see also Appendix 29 in Ref. \cite{ZJ}.



\end{thebibliography}
\end{document}